# Transducer finite aperture effects in sound transmission near leaky Lamb modes in elastic plates at normal incidence

Magne Aanes[1,2], Kjetil Daae Lohne[2], Per Lunde[1,2], Magne Vestrheim[1]

[1] University of Bergen, Department of Physics and Technology,
P.O. Box 7803, N-5020 Bergen, Norway
[2] Christian Michelsen Research AS (CMR),
P.O. Box 6031 Postterminalen, N-5892 Bergen, Norway
Contact email: magne.aanes@uib.no

## Abstract

The interaction of ultrasonic waves with fluid-embedded viscoelastic plates, pipes, and shells, have been subject to extensive theoretical and experimental studies over several decades. In normal-incidence through-transmission measurements of a water-embedded solid plate using ultrasonic piezoelectric transducer sound fields, significant deviations from plane wave theory have recently been observed. To quantitatively describe such measured phenomena, finite element modeling (FEM), also combined with an angular spectrum method (ASM), have been used for three-dimensional (3D) simulation of the voltage-to-sound-pressure signal propagation through the electro-acoustic measurement system consisting of the piezoelectric transducer, the water-embedded steel plate, and the fluid regions at both sides of the plate. The observed phenomena of frequency downshift of the plate resonance, increased sound pressure level through the plate, and beam narrowing / widening, are ascribed to the finite angular spectrum of the beam, that excites a region of negative and zero group velocity for the leaky Lamb mode in question.

**SUMMARY:** The interaction of ultrasonic waves with fluid-embedded viscoelastic plates, pipes, and shells, have been subject to extensive theoretical and experimental studies over several decades. Ultrasonic guided waves (GUW) represent important scientific and industrial fields [1-26]. Examples include non-destructive testing and evaluation, measurement of elastic properties of solid materials, wall thickness measurement and corrosion measurement in pipes, deposit detection, integrity evaluation, and fluid flow measurement. In such applications the plane-wave theory of fluid-embedded viscoelastic plates is often used, in design of measurement methods as well as for signal interpretation [1-4].

In normal-incidence through-transmission measurements of a water-embedded solid plate using ultrasonic piezoelectric transducer sound fields, significant deviations from plane wave theory have been observed. Downward frequency shift for some plate resonances and enhanced signal transmission as compared to plane-wave theory have been demonstrated in measurements and simulations [15-19]. These phenomena may be accompanied by beam narrowing (and in other cases beam widening) in the transmitted field [18,19]. Such observations are made e.g. in the frequency region of the important

   1

fundamental thickness-extensional (TE) mode of a steel plate, corresponding to the cutoff frequency of the S<sub>1</sub> Lamb mode.fundamental thickness-extensional (TE) mode of a steel plate, corresponding to the cutoff frequency of the $S_1$ Lamb mode.[1]

To quantitatively describe such measured phenomena, finite element modeling (FEM), also combined with an angular spectrum method (ASM), have been used for three-dimensional (3D) simulation of the voltage-to-sound-pressure signal propagation through the electro-acoustic measurement system consisting of the piezoelectric transducer, the water-embedded steel plate, and the fluid regions at both sides of the plate [15-19]. Theoretical and measured results show that the transmission through the plate can be studied quantitatively for excitation of leaky Lamb modes in the plate using real acoustic beams. For more generic studies, to avoid relying on the particular transducer being used in these measurements, a simplified model has also been used, in which the 3D transducer sound field incident to the plate is approximated by the far field radiated by a baffled and uniformly vibrating, circular and planar piston source. In earlier works fair agreement has been obtained on the acoustic axis in comparison of this model with measurements and the more complete simulation models [15-19].

In the present work, the latter model is used to investigate the observed deviations between measurements and plane-wave theory, by systematically varying the Poisson's ratio in the plate, in a frequency band covering the lower leaky Lamb modes of the plate. A 21.1 mm diameter piston transducer is radiating at normal beam incidence to a $d$ = 6.05 mm thick water-embedded solid plate, at a distance $z_0$ = 270 mm from the plate. At 478 kHz the *ka* number is about 21, and the Rayleigh distance is about 112 mm, well below $z_0$. The Poisson's ratio of the plate, σ, is varied over the range 0 - 0.5 by setting the plate's compressional-wave velocity to a fixed value, $c_\ell$ = 5780 m/s (corresponding to steel), and varying the shear-wave velocity, $c_t$ [19]. For this incident beam field, the sound pressure field transmitted through the plate is studied as a function of the Poisson's ratio, σ, and the signal frequency, *f*. This includes the pressure-to-pressure transfer function of the plate, $H_{PP}^{plate}$, defined as the ratio of the transmitted axial sound pressure at the lower surface, to the incident free-field axial sound pressure at the upper surface of the plate, at the frequency *f*.

As an example, Fig. 1a shows $|H_{PP}^{plate}|$ calculated over the frequency band 382 to 535 kHz, for Poisson's ratio in the range 0 - 0.5. The frequency axis is normalized to the fundamental TE mode of the corresponding vacuum-embedded plate, $c_\ell/2d \approx$ 478 kHz. Note that only a limited frequency range is shown in this example, covering the range of the fundamental TE mode. The plane-wave cutoff frequencies for Lamb modes of the corresponding vacuum-embedded plate, $f_c$, are shown using white curves. The white vertical line designates the fundamental TE mode, and non-vertical white curves designate thickness-shear (TS) modes. Solid and dotted curves represent symmetric (S) and anti-symmetric (A) modes, respectively. The black curve with white circled markers gives $f_0$, the zero-group velocity (ZGV) frequency of the $S_1$ Lamb mode in the vacuum-embedded plate. From the figure, it is observed that for all σ, a frequency of maximum $|H_{PP}^{plate}|$ (i.e., maximum beam transmission), denoted $f_b$, exists in-between the ZGV frequency, $f_0$, and the nearest cutoff frequency, $f_c$. Over the σ region 0 - 0.5, resonance frequency downshifts in the range of approximately 0.92 to 0.98 may be experienced, relative to the nearest $f_c$. The frequency downshift appears to be largest for σ close to 1/3.

Further details are given in Fig. 1b, for a steel plate with σ = 0.2925 (AISI 316L stainless steel). $|H_{PP}^{plate}|$ is shown as a function of frequency (red curve), in comparison with the magnitude of the corresponding plane-wave pressure transmission coefficient for a plane-wave at 1º incidence (blue curve). As in Fig. 1a, it is noted that $f_b$ is located in the region between $f_0$ and $f_c$, at about $0.95 f_c$. In addition, an enhanced axial signal transmission through the plate of about +4.1 dB is observed at $f_b$. This means that for the finite beam in question, at the frequency $f_b$, the axial sound pressure level (SPL) is about 4.1 dB higher at the plate's lower surface than at the upper surface.

---

[1] Here, Lamb modes are numbered according to increasing cutoff frequency, and "S" denotes a symmetric mode

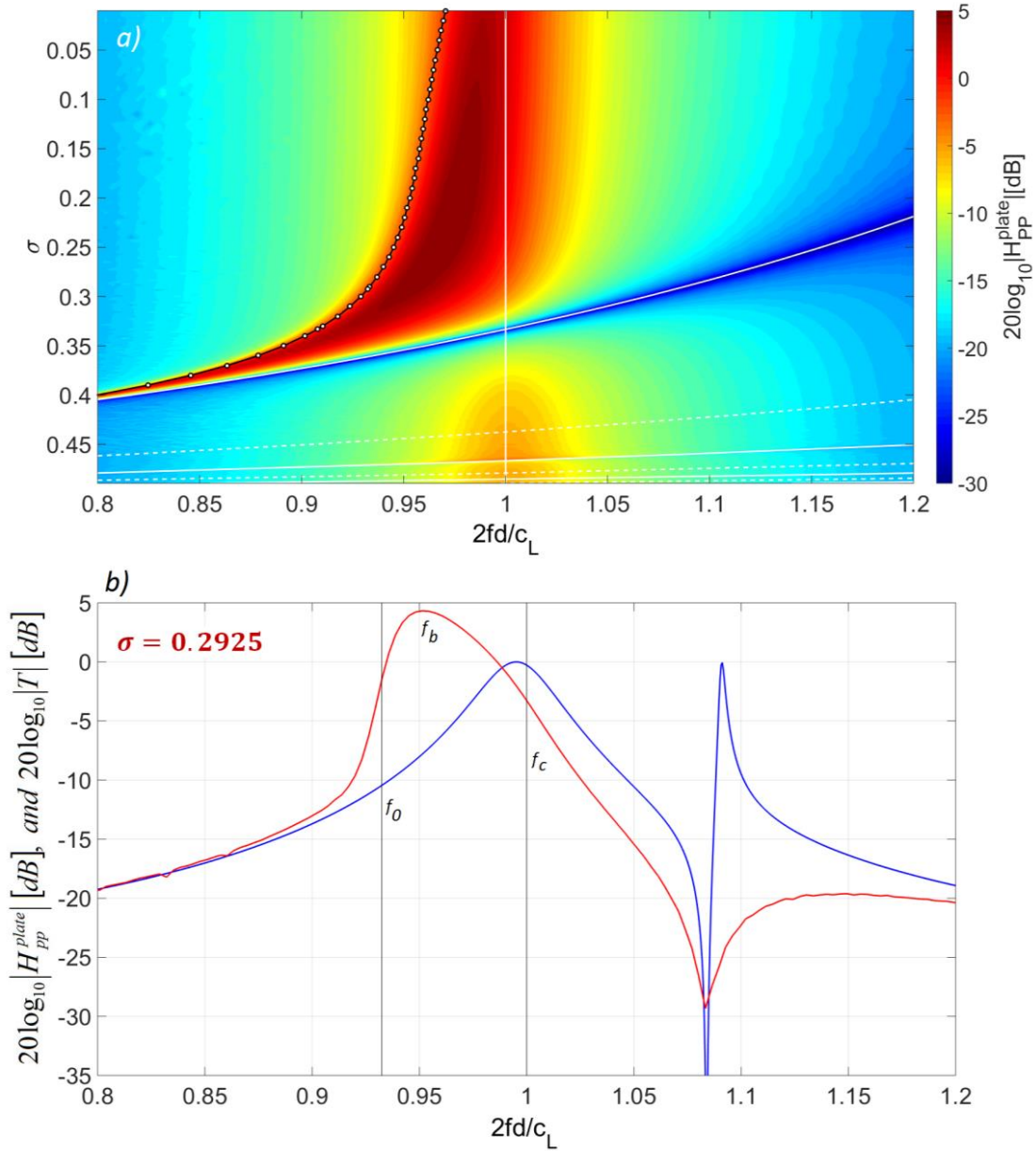

Fig. 1. (a) The magnitude of the pressure-pressure transfer function for beam transmission through a water-embedded plate, $|H_{PP}^{plate}|$, shown as a function of the normalized frequency and the Poisson's ratio in the plate, over a limited frequency span. The cutoff frequencies of Lamb modes in the associated vacuum-embedded plate, $f_c$, are shown using white curves. The ZGV frequency of the $S_1$ Lamb mode, $f_0$, is shown using a black curve with white circled markers.

(b) $|H_{PP}^{plate}|$ for beam transmission through a steel plate with Poisson's ratio equal to 0.2925 (red curve), compared with the plane-wave pressure transmission coefficient of the plate, $|T|$, for a plane-wave at $1^0$ incidence (blue curve). The frequencies $f_0$, $f_b$ and $f_c$ of the $S_1$ Lamb mode are indicated.

The observed phenomena of frequency downshift of the plate resonance, increased sound pressure level through the plate, and beam narrowing / widening (not illustrated in the example shown here), are ascribed to the finite angular spectrum of the beam, that excites a region of negative and zero group velocity for the leaky Lamb mode in question [19]. The finite beam, combined with the negative group velocity (NGV) region, appears to result in a concentration of the beam energy in the plate, close to the acoustic axis. The plate is put into resonance at a lower frequency than the Lamb mode's cutoff frequency, with re-radiation of sound into the fluid at a higher axial SPL than that of the incident field. The phenomena addressed here may have important influences e.g. for ultrasonic methods for material characterization and thickness measurement.

In the literature, regions of negative group velocity for certain Lamb modes have been associated with backward wave propagation, phase and group velocities of opposite sign, and energy transport in the negative direction [20-26]. The apparent correlation between such Lamb wave phenomena discussed in the literature, and the finite beam diffraction Lamb wave phenomena described here, will be addressed in further work.

# References

[1] B. Hosten and M. Castaings, "Transfer matrix of multilayered absorbing and anisotropic media. Measurements and simulations of ultrasonic wave propagation through composite materials", J. Acoust. Soc. Am. **94**, 1488-95 (1993).

[2] P. Cawley and B. Hosten, "The use of large ultrasonic transducers to improve transmission coefficient measurements on viscoelastic anisotropic plates", J. Acoust. Soc. Am. **101**, 1373-79 (1997).

[3] D. Fei, D.E. Chimenti, and S.V. Teles, "Material property estimation in thin plates using focused, synthetic-aperture acoustic beams", J. Acoust. Soc. Am. **113**(5), 2599-2610 (2003).

[4] J. Jocker and D. Smeulders, "Minimization of finite beam effects in the determination of reflection and transmission coefficients of an elastic layer", Ultrasonics, **46**, 42-50 (2007).

[5] A. Gibson and J.S. Popovics, "Lamb wave basis for impact-echo method analysis", J. Eng. Mech, 438-443 (April 2005).

[6] Y.T. Tsai and J. Zhu, "Simulation and experiments of airborne zero-group-velocity Lamb waves in concrete plate", J. Nondestr. Eval. **31**, 373-382 (2012).

[7] R.K. Johnson and A.J. Devaney, "Transducer model for plate thickness measurement", Proc. Ultrason. Symp., San Diego, CA (1982), pp. 502-504.

[8] M. Castaings and P. Cawley, "The generation, propagation, and detection of Lamb waves in plates using air-coupled ultrasonic transducers," J. Acoust. Soc. Am., **100**(5), 3070-3077 (1996).

[9] E. Moulin, J. Assaad, C. Delebarre, M. Houy, V. Cedex, and D. Osmont, "Modeling of Lamb waves generated by integrated transducers in composite plates using a coupled finite element-normal modes expansion method," J. Acoust. Soc. Am., **107**(1), 87-94 (2000).

[10] M. Bezdêk and B. R. Tittmann, "Dispersion analysis of a three-layered waveguide with finite element and matrix methods," Acta Acustica united with Acustica, **94**(5), 792-806 (2008).

[11] B. Hosten and C. Biateau, "Finite element simulation of the generation and detection by air-coupled transducers of guided waves in viscoelastic and anisotropic materials", J. Acoust. Soc. Am. **123**(4), 1963-71 (2008).

[12] W. Ke, M. Castaings and C. Bacon, "3D finite element simulations of an air-coupled ultrasonic NDT system", NDT&E Intern. **42**(6), 524-533 (2009).

[13] S. Delrue, K. van den Abeele, E. Blomme, J. Deveugele, P. Lust, and O.B. Matar, "Two-dimensional simulation of the single-sided air-coupled ultrasonic pitch-catch technique for non-destructive testing," Ultrasonics **50**(2), 188-196 (2010).

[14] M. Masmoudi, B. Hosten, and C. Biateau, "Analytical and finite element methods for studying the influence of the air-coupled transducer characteristics on the purity of guided waves generated in solids," Rev. Quantit. Nondestr. Eval. **29**, 1887-94 (2010).

[15] K.D. Lohne, P. Lunde, and M. Vestrheim, "Measurements and 3D simulations of ultrasonic directive beam transmission through a water-immersed steel plate", Proc. 34th Scand. Symp. Phys. Acoust., Geilo, Norway, 30 Jan. - 2 Feb., 2011.

[16] M. Aanes, K.D. Lohne, P. Lunde, and M. Vestrheim, "Normal incidence ultrasonic transmission through a water-immersed plate using a piezoelectric transducer. Finite element modeling, angular spectrum method, and measurements", Proc. Int. Congr. Sound Vib., Vilnius, Lithuania, 8-12 July 2012.

[17] M. Aanes, K.D. Lohne, P. Lunde, and M. Vestrheim, "Ultrasonic beam transmission through a water-immersed plate at oblique incidence using a piezoelectric source transducer. Finite element - angular spectrum modeling and measurements", Proc. IEEE Int. Ultras. Symp., Dresden, Germany, 7-10 Oct. 2012, pp. 1972-77.

[18] M. Aanes, "Interaction of piezoelectric transducer excited ultrasonic pulsed beams with a fluid-embedded viscoelastic plate", PhD thesis, Dept. of Physics and Technology, Univ. of Bergen, Bergen, Norway (2013).

[19] M. Aanes, K.D. Lohne, P. Lunde, and M. Vestrheim, "Transducer beam diffraction effects in sound transmission near leaky Lamb modes in elastic plates at normal incidence," in *Proc. 2015 IEEE Int. Ultras. Symp.*, Taipei, Taiwan, 21-24 Oct. 2015, pp. 1-4.

[20] I. Tolstoy and E. Usdin, Wave propagation in elastic plates: Low and high mode dispersion, J. Acoust. Soc. Am., **29**(1), 37-42 (1957).

[21] A.H. Meitzler, "Backward-wave transmission of stress pulses in elastic cylinders and plates," J. Acoust. Soc. Am., **38**, 835-842 (1965).

[22] J. Wolf, T.D.K. Ngoc, R. Kille, and W.G. Mayer, "Investigation of Lamb waves having negative group velocity", J. Acoust. Soc. Am. **83**, 122-126 (1988).

[23] M.F. Werby, and H. Überall, "The analysis and interpretation of some special properties of higher order symmetric Lamb waves: The case for plates", J. Acoust. Soc. Am. 111, 2686-2691 (2002).

[24] S.D. Holland and D.E. Chimenti, "Air-coupled acoustic imaging with zero-group velocity Lamb modes", Appl. Phys. Lett. **83**(13), 2704-06 (2003).

[25] C. Prada, O. Balogun, and T.W. Murray, "Laser-based" ultrasonic generation and detection of zero-group velocity Lamb waves in thin plates", Appl. Phys. Lett. **87**, 194109 (2005).

[26] C. Prada, D. Clorennec, and D. Royer, "Local vibration of an elastic plate and zero-group velocity Lamb modes", J. Acoust. Soc. Am. **124**(1), 203-212 (2008).